\documentclass[%
 reprint,
 amsmath,amssymb,
 aps,
]{revtex4-2}

\usepackage{graphicx}
\usepackage{dcolumn}
\usepackage{bm}

\usepackage{chemformula} 
\usepackage[version = 4]{mhchem} 
\usepackage[algoruled, linesnumbered]{algorithm2e}
\usepackage{braket}

\begin{document}

\title{From Ground State to Electronic and Vibronic Spectra \textit{via} Memory Kernel Coupling Theory and Neural Quantum States}

\author{Xinchen He}
\altaffiliation{Department of Chemistry, Westlake University, Hangzhou, Zhejiang 310030, China}
\altaffiliation{Institute of Natural Sciences, Westlake Institute for Advanced Study, Hangzhou 310030, China}
\altaffiliation{Current Address: Department of Chemistry, Molecular Science Research Hub (MSRH), Imperial College London, White City Campus, 82 Wood Lane, London W12 0BZ, UK}

\author{Wenjie Dou}
\email{douwenjie@westlake.edu.cn}
\altaffiliation{Department of Chemistry, Westlake University, Hangzhou, Zhejiang 310030, China}
\altaffiliation{Department of Physics, Westlake University, Hangzhou, Zhejiang 310030, China}
\altaffiliation{Institute of Natural Sciences, Westlake Institute for Advanced Study, Hangzhou 310030, China}

\date{\today}

\begin{abstract}
Modelling molecular absorption spectra from scalable ground-state representations is challenging because conventional approaches often require explicit excited states or time propagation. Here, we introduce a method to obtain Absorption Using Recursive Operator Response with vibronic Augmentation (Aurora), which combines neural quantum states (NQSs) with continued-fraction memory-kernel coupling theory (CF-MKCT). Aurora maps the zero-temperature Liouvillian response hierarchy onto repeated wavefunction-space applications of a shifted Hamiltonian to a dipole-prepared state, enabling direct spectral reconstruction from variational ground states. The implementation combines variational Monte Carlo, implicit-restarted-Lanczos-assisted NQS optimization, matrix-free Hamiltonian application, and adaptive recursion control. The resulting electronic spectra closely agree with full-configuration-interaction and conventional MKCT references for tractable molecules, with the restarted-Lanczos strategy improving difficult NQS cases. For electron-phonon Hamiltonians, Aurora combines Huang-Rhys mode selection with local singular-value-decomposition (SVD) phonon encoders. The \ce{H2} benchmark indicates that compression retains the dominant vibronic envelope and improves upon bare local-basis truncation, while paired calculations for \ce{N2}, \ce{H2O}, \ce{H2S}, and \ce{NH3} yield similar principal features with and without encoding. We further obtain vibronic spectra for \ce{C6H6} and a 10-electron, 10-orbital \ce{C10H8} active space. Overall, Aurora provides a unified route from variational molecular ground states to electronic and vibronic spectra without constructing excited-state manifolds.
\end{abstract}

\maketitle


\section{\label{sec:level1}Introduction}
The Born-Oppenheimer (BO) approximation plays a foundational role in molecular quantum mechanics. By separating electronic and nuclear motion, BO approximation reduces the molecular problem to electronic-structure calculations at fixed nuclear configurations.\textsuperscript{\cite{bornZurQuantentheorieMolekeln1927}} This separation enables efficient descriptions of molecular structure, electronic excitations, and spectroscopy. However, this approximation is not universally adequate. Electronic and nuclear motion become inseparable near avoided crossings and conical intersections, or when vibronic and electron--phonon couplings are strong.\textsuperscript{\cite{tullyMolecularDynamicsElectronic1990, giustinoElectronphononInteractionsFirst2017, lehrRoleVibronicCoupling2020}} Retaining these coupled degrees of freedom causes the joint Hilbert space to grow rapidly with the electronic configurations, vibrational modes, and local phonon states.\textsuperscript{\cite{baiardiGeneralTimeDependent2013,lehrRoleVibronicCoupling2020,giustinoElectronphononInteractionsFirst2017}} Methods such as multi-configurational time-dependent Hartree approach (MCTDH), Multilayer formulation of the multiconfiguration time-dependent Hartree theory (ML-MCTDH), and time-dependent density matrix renormalization group (TD-DMRG) mitigate this growth through adaptive wavefunction representations to some extent, but their construction of spectra commonly relies on explicit real-time propagation.\textsuperscript{\cite{meyerMulticonfigurationalTimedependentHartree1990, wangMultilayerFormulationMulticonfiguration2003, renTimeDependentDensityMatrix2018}}

An effective electron-phonon Hamiltonian provides a route beyond the BO picture. In this representation, vibrational modes are quantized, coupled to the electronic Hamiltonian, and included in a joint electron-phonon wavefunction. The resulting ground state naturally contains electron-phonon coupling rather than factorizing into independent electronic and vibrational states. Here, the vibrational modes are harmonic and their interaction with the electrons is described by a linear-coupling model.\textsuperscript{\cite{giustinoElectronphononInteractionsFirst2017}} The corresponding Huang-Rhys factors quantify the coupling associated with individual modes.\textsuperscript{\cite{weiPolaritonicHuangRhys2023}} To control the growth of the joint Hilbert space, we retain modes according to a relative Huang--Rhys criterion. Variational Monte Carlo methods can then represent the resulting many-body ground state using neural quantum states (NQSs).\textsuperscript{\cite{chooFermionicNeuralnetworkStates2020, pfauInitioSolutionManyelectron2020, carleoSolvingQuantumManybody2017}}

The NQS ground-state ansatz provides a natural starting point for spectroscopy modelling. For an excitation operator $\hat{\mu}$, the zero-temperature response is determined by the Hamiltonian, the ground state, and the dipole-prepared state $\hat{\mu}\lvert\Psi_0\rangle$.\textsuperscript{\cite{baiardiGeneralTimeDependent2013}} Repeated applications of the Hamiltonian to this state generate a hierarchy of quantities that encodes transitions to the accessible excited-state manifold. Our previous kernel-polynomial NQS (KPNQS) approach exploited this hierarchy through a Chebyshev recursion of spectral moments.\textsuperscript{\cite{liuAbsorptionSpectraKernel2025}} However, producing well-resolved electronic absorption spectra required on the order of $10^4$ recursion steps \textit{via} KPNQS.\textsuperscript{\cite{liuAbsorptionSpectraKernel2025}}

This motivates us to develop a more compact formulation to obtain excited-state properties from ground state character. Memory-kernel coupling theory (MKCT) represents dynamical response through a recursive hierarchy of coupled memory kernels, thereby avoiding explicit real-time propagation.\textsuperscript{\cite{liuMemoryKernelCoupling2025,biGeneralizedQuantumMaster2026}} Its continued-fraction formulation, CF-MKCT, reorganizes this hierarchy into a compact frequency-domain representation.\textsuperscript{\cite{liuStableMemoryKernel2026}} This construction offers a potential route to reduce the recursion depth of KPNQS while preserving its avoidance of explicit excited-state construction and real-time propagation. In model open quantum systems, CF-MKCT has reproduced dissipaton equation-of-motion \textsuperscript{\cite{tanimuraNumericallyExactApproach2020,yanDissipationEquationMotion2016, wangDissipatonDynamicsTheory2018}} benchmarks to numerical accuracy, supporting the accuracy and stability of the continued-fraction hierarchy.\textsuperscript{\cite{liuStableMemoryKernel2026}} However, CF-MKCT has remained primarily an operator-space method for model systems. A wavefunction-space formulation compatible with variational molecular ground states, particularly joint electron--phonon states, has not been established.

Here we introduce a new way to obtain Absorption Using Recursive Operator Response with vibronic Augmentation (Aurora), an NQS-integrated CF-MKCT framework for molecular absorption spectra. We first establish a zero-temperature wavefunction-space formulation of CF-MKCT. For a pure ground-state density operator, we map the Liouvillian recursion onto repeated applications of a shifted Hamiltonian to a normalized dipole-prepared state. This operator--wavefunction correspondence converts the memory-kernel hierarchy into quantities that can be evaluated directly using a variational wavefunction backend. We realize this formulation using an NQS representation of the molecular ground state, without constructing the full excited-state manifold. We further extend the framework to electron--phonon Hamiltonians by integrating Huang--Rhys mode selection, qubit representations, singular-value-decomposition-based phonon compression, and CF-MKCT spectral reconstruction.

Aurora builds on and advances both parent lines of work. Compared with the previous KPNQS approach, Aurora replaces the long Chebyshev recursion with a continued-fraction formulation that reaches spectral convergence with substantially fewer iterations. This reduction is especially important for electron--phonon-coupled systems, whose joint Hilbert spaces grow rapidly. Compared with earlier CF-MKCT applications, Aurora provides a molecular electronic-structure interface to NQS ground states, electronic and vibronic Hamiltonians, qubit representations of electronic and phonon degrees of freedom, and spectral convergence diagnostics. Together, these developments extend CF-MKCT beyond model open quantum systems and establish it as a practical framework for molecular spectroscopy. From a conceptual perspective, this method enables the calculation of dynamical properties without invoking the Born–Oppenheimer (BO) approximation.

We evaluate Aurora in two stages. First, we benchmark electronic absorption spectra against full-configuration-interaction (FCI) and conventional MKCT references, testing the recursive construction and NQS ground states. Second, we examine vibronic spectra with and without phonon encoding across systems ranging from tractable molecules to larger aromatic compounds. These benchmarks assess spectral convergence, NQS accuracy, and phonon compression as the accessible Hilbert space increases.

\section{Theory and Methodology}
\label{sec:method}
The central objective of the present framework is to evaluate zero-temperature absorption spectra from a ground-state wavefunction representation while retaining the recursive structure of memory-kernel coupling theory (MKCT). We first establish the operator-wavefunction correspondence underlying this construction, then
describe its realization with electronic neural quantum states (NQSs), and finally extend the same formalism to explicitly coupled electron-phonon Hamiltonians.

\subsection{Electronic Hamiltonian and optical response}
\label{sec:electronic_problem}

With a frozen-core approximation,
the second-quantized Hamiltonian is\textsuperscript{\cite{mcardleQuantumComputationalChemistry2020}}
\begin{equation}
\begin{aligned}
\hat H_{\mathrm{el}}
&=
E_{\mathrm{nuc}}+E_{\mathrm{core}}\\
&+\sum_{pq} h_{pq}\hat a_p^\dagger \hat a_q
+\frac{1}{2}\sum_{pqrs} h_{pqrs}
\hat a_p^\dagger \hat a_q^\dagger \hat a_r \hat a_s
\end{aligned}
\label{eq:electronic_hamiltonian}
\end{equation}
where \(h_{pq}\) and \(h_{pqrs}\) denote the one- and two-electron integrals, and \(\hat a_p^\dagger\) (\(\hat a_p\)) creates (annihilates) an
electron in spin orbital \(p\). Molecular geometries, electronic integrals, and
reference wavefunctions are generated with PySCF \textsuperscript{\cite{sunPythonSimulationsChemistry2026,sunRecentDevelopmentsSCF2020,sunSCFPythonbasedSimulations2018,sunLibcintEfficientGeneral2015}}. Unless otherwise stated, STO-3G is used for the electronic basis in the benchmark
calculations. The frozen-core spaces used in the calculations are
summarized in Table~\ref{tab:active_space}.

\begin{table}[h]
\centering
\caption{Frozen-core active spaces used in the benchmark calculations.}
\begin{tabular}{lcccc}
\hline
& \multicolumn{2}{c}{Total} & \multicolumn{2}{c}{Active} \\
\cline{2-3}\cline{4-5}
Molecule & \(N_e\) & \(N_o\) & Electrons & Orbitals \\
\hline
\ce{H2O}   & 10 & 7  & 8  & 6  \\
\ce{NH3}   & 10 & 8  & 8  & 7  \\
\ce{C2}    & 12 & 10 & 8  & 8  \\
\ce{N2}    & 14 & 10 & 10 & 8 \\
\ce{O2}    & 16 & 10 & 12 & 8 \\
\ce{C6H6} & 42 & 36 & 6 & 6 \\
\ce{C10H8} & 68 & 58 & 10 & 10 \\
\hline
\end{tabular}
\label{tab:active_space}
\end{table}

For polarization mode \(\alpha\), the electronic dipole operator is
\begin{equation}
\begin{aligned}
\hat\mu_\alpha
&=
\sum_{pq}\mu_{pq}^{(\alpha)}
\hat a_p^\dagger \hat a_q
\end{aligned}
\label{eq:dipole_operator}
\end{equation}
To remove the zero-frequency contribution associated with the ground-state
permanent dipole, we use the centred dipole operator
\begin{equation}
\begin{aligned}
\delta\hat\mu_\alpha
&:=
\hat\mu_\alpha
-
\langle\Psi_0|\hat\mu_\alpha|\Psi_0\rangle \hat I
\end{aligned}
\label{eq:centered_dipole}
\end{equation}
The corresponding dipole-prepared state is calculated as
\begin{equation}
\begin{aligned}
|\phi_\alpha\rangle
&:=
\delta\hat\mu_\alpha|\Psi_0\rangle
\end{aligned}
\label{eq:dipole_prepared_state}
\end{equation}

\subsection{Zero-temperature MKCT in wavefunction space}
\label{sec:mkct_zero_temperature}

\subsubsection{Autocorrelation and Shifted Hamiltonian}

\noindent At zero temperature, the dipole-dipole correlation function can be written as
\begin{equation}
\begin{aligned}
C_\alpha(t)
&=
\langle\Psi_0|
\delta\hat\mu_\alpha(t)
\delta\hat\mu_\alpha
|\Psi_0\rangle \\
&=
\langle\phi_\alpha|
e^{-i(\hat H-E_0)t}
|\phi_\alpha\rangle
\end{aligned}
\label{eq:dipole_autocorrelation}
\end{equation}
We therefore define the shifted Hamiltonian
\begin{equation}
\begin{aligned}
\hat G
&:=
\hat H-E_0\hat I
\end{aligned}
\label{eq:shifted_hamiltonian}
\end{equation}
and the normalized correlation function
\begin{equation}
\begin{aligned}
c_\alpha(t)
&:=
\frac{C_\alpha(t)}{C_\alpha(0)}
\qquad
c_\alpha(0)=1
\end{aligned}
\label{eq:normalized_autocorrelation}
\end{equation}

\subsubsection{Operator--wavefunction correspondence}

MKCT is naturally formulated in operator space.\textsuperscript{\cite{biGeneralizedQuantumMaster2026}} For the zero-temperature density
operator \(\rho_0=|\Psi_0\rangle\langle\Psi_0|\), its inner product reduces to
\begin{equation}
\begin{aligned}
(\hat O_1,\hat O_2)_0
&:=
\operatorname{Tr}
\!\left[
\hat O_1^\dagger\hat O_2\rho_0
\right] \\
&=
\langle\Psi_0|
\hat O_1^\dagger\hat O_2
|\Psi_0\rangle
\end{aligned}
\label{eq:operator_inner_product}
\end{equation}
This motivates the map
\begin{equation}
\begin{aligned}
\mathcal J:\hat O
&\mapsto
\hat O|\Psi_0\rangle
\end{aligned}
\label{eq:J_map}
\end{equation}
which preserves the inner product on the subspace generated from
\(|\Psi_0\rangle\):
\begin{equation}
\begin{aligned}
(\hat O_1,\hat O_2)_0
&=
\langle
\mathcal J(\hat O_1)
|
\mathcal J(\hat O_2)
\rangle .
\end{aligned}
\label{eq:J_isometry}
\end{equation}

\noindent We can map the Liouvillian Opterator
\(\mathcal L\hat O=[\hat H,\hat O]\) as
\begin{equation}
\begin{aligned}
\mathcal J(\mathcal L\hat O)
&=
[\hat H,\hat O]|\Psi_0\rangle \\
&=
(\hat H-E_0\hat I)\hat O|\Psi_0\rangle \\
&=
\hat G\,\mathcal J(\hat O).
\end{aligned}
\label{eq:liouvillian_map}
\end{equation}
and multiple application of Liouvillian can be mapped as,
\begin{equation}
\begin{aligned}
\mathcal J(\mathcal L^n\hat O)
&=
\hat G^n\mathcal J(\hat O)
\end{aligned}
\label{eq:liouvillian_power_map}
\end{equation}
Therefore, the zero-temperature MKCT hierarchy generated in operator space ($\mathcal K^{op}_N$) can be mapped with repeated action of \(\hat G\) on a wavefunction in space $\mathcal K^\text{wfn}_N$ using $\mathcal J$.
\begin{equation}
\mathcal K_N^{\mathrm{op}}
:=
\operatorname{span}
\left\{
\hat A,
\mathcal L\hat A,
\ldots,
\mathcal L^{N-1}\hat A
\right\}
\end{equation}
\begin{equation}
\mathcal K_N^{\mathrm{wf}}
:=
\operatorname{span}
\left\{
|q_0\rangle,
\hat G|q_0\rangle,
\ldots,
\hat G^{N-1}|q_0\rangle
\right\}
\end{equation}
\begin{equation}
\mathcal J
\left(
\mathcal K_N^{\mathrm{op}}
\right)
=
\mathcal K_N^{\mathrm{wf}}
\label{eqn:central_map}
\end{equation}
with $\lvert q_0 \rangle $ defined as
\begin{equation}
\lvert q_0 \rangle :=
\frac{\mathcal J(\hat A)}{\|\mathcal J(\hat A)\|}=
\frac{\lvert \phi_a \rangle}{\sqrt{C_{0,a}}}
\label{eqn: Lanczos_init}
\end{equation}

Equation \ref{eqn:central_map} provides the central zero-temperature map used in this work. It closes the MKCT Liouvillian basis to the wavefunction space, allowing successive recursion orders to be constructed without introducing an explicit Liouville-space representation and thereby providing a direct interface to variational many-body states.

\subsubsection{Moment equivalence and recursive spectral representation}
\label{sec: cf-spectral-gen}
For \(\hat A=\delta\hat\mu_\alpha\), define the real spectral moments of the shifted Hamiltonian as
\begin{equation}
\begin{aligned}
m_n^{(\alpha)}
&:=
\frac{
\langle\phi_\alpha|
\hat G^n
|\phi_\alpha\rangle
}{
\langle\phi_\alpha|\phi_\alpha\rangle
}
\end{aligned}
\label{eq:wavefunction_moments}
\end{equation}
The corresponding MKCT moments,
\begin{equation}
\begin{aligned}
\Omega_n^{(\alpha)}
&:=
\frac{
\bigl((i\mathcal L)^n\hat A,\hat A\bigr)_0
}{
(\hat A,\hat A)_0
}
\end{aligned}
\label{eq:mkct_moments}
\end{equation}
obey
\begin{equation}
\begin{aligned}
m_n^{(\alpha)} = \mathcal J(\Omega_n)=i^{-n}\Omega_n
\end{aligned}
\label{eq:moment_equivalence}
\end{equation}
Thus the information entering the MKCT hierarchy is recoverable from moments of
\(\hat G\) evaluated in the dipole-prepared state.

\noindent A numerically stable realization of the same recursion is obtained by
tridiagonalizing \(\hat G\) from the normalized initial vector
\begin{equation}
\begin{aligned}
|q_0\rangle
&:=
\frac{|\phi_\alpha\rangle}
{\sqrt{C_\alpha(0)}} .
\end{aligned}
\label{eq:q0}
\end{equation}
The three-term recurrence is\textsuperscript{\cite{dargelAdaptiveLanczosvectorMethod2011, giannozziOrdinaryMatrixContinued1988}}
\begin{equation}
\begin{aligned}
\alpha_n
&=
\langle q_n|\hat G|q_n\rangle,\\
|r_n\rangle
&=
\hat G|q_n\rangle
-\alpha_n|q_n\rangle
-\beta_{n-1}|q_{n-1}\rangle,\\
\beta_n
&=
\|r_n\|,\\
|q_{n+1}\rangle
&=
\frac{|r_n\rangle}{\beta_n}
\end{aligned}
\label{eq:lanczos_recurrence}
\end{equation}
with the Jacobi matrix
\begin{equation}
\begin{aligned}
T_N
&=
\begin{pmatrix}
\alpha_0 & \beta_0 & 0 & \cdots\\
\beta_0 & \alpha_1 & \beta_1 & \ddots\\
0 & \beta_1 & \alpha_2 & \ddots\\
\vdots & \ddots & \ddots & \ddots
\end{pmatrix}
\end{aligned}
\label{eq:jacobi_matrix}
\end{equation}

The tridiagonalization is used here as a stable numerical representation of the zero-temperature MKCT recursion rather than as an independent spectral approximation. In the notation of the continued-fraction MKCT hierarchy\textsuperscript{\cite{liuStableMemoryKernel2026}}, the coefficients are related by
\begin{equation}
\begin{aligned}
A_n&=-i\alpha_n,
&
B_n&=\beta_n^2
\end{aligned}
\label{eq:AB_map}
\end{equation}
Hence the Laplace-transformed normalized correlation function takes the finite
continued-fraction form
\begin{equation}
\begin{aligned}
\widetilde c_\alpha^{[N]}(s)
&=
\cfrac{1}{
s+i\alpha_0+
\cfrac{\beta_0^2}{
s+i\alpha_1+
\cfrac{\beta_1^2}{
s+i\alpha_2+\ddots
}
}
}
\end{aligned}
\label{eq:continued_fraction}
\end{equation}

\noindent Diagonalizing $T_N=U\Omega U^\dagger$ would obtain
$\Omega=\operatorname{diag}(\omega_1,\ldots,\omega_N)$, gives\textsuperscript{\cite{dargelAdaptiveLanczosvectorMethod2011}}
\begin{equation}
\begin{aligned}
\widetilde c_\alpha^{[N]}(s)
&=
\sum_{k=1}^{N}
\frac{|U_{0k}|^2}{s+i\omega_k}
\end{aligned}
\label{eq:ritz_correlation}
\end{equation}
The corresponding spectral weights are
\begin{equation}
\begin{aligned}
w_k^{(\alpha)}
&=
C_\alpha(0)|U_{0k}|^2 
\end{aligned}
\label{eq:stick_weights}
\end{equation}
For visualization, we broaden the discrete spectrum with a normalized Gaussian,
\begin{equation}
\begin{aligned}
I_\alpha(\omega)
&=
2\pi
\sum_k
w_k^{(\alpha)}
\frac{1}{\sqrt{2\pi}\sigma}
\exp\!\left[
-\frac{(\omega-\omega_k)^2}{2\sigma^2}
\right]
\end{aligned}
\label{eq:broadened_spectrum}
\end{equation}
While the continued fraction is unlikely to be finite, we use the following autoconvergence procedure for spectral construction. Finite-precision arithmetic can lead to loss of orthogonality in a long
three-term recurrence. We monitor the maximum and mean deviations of the Gram
matrix from the identity,
\begin{equation}
\begin{aligned}
\mathcal L_{\max}
&=
\max_{0\le i,j<N}
\left|
\langle q_i|q_j\rangle-\delta_{ij}
\right|
\\
\mathcal L_{\mathrm{mean}}
&=
\frac{1}{N^2}
\sum_{i,j=0}^{N-1}
\left|
\langle q_i|q_j\rangle-\delta_{ij}
\right|
\end{aligned}
\label{eq:orthogonality_metrics}
\end{equation}
When required, the residual vector is reorthogonalized against the existing basis,
\begin{equation}
\begin{aligned}
|r_n\rangle
&\leftarrow
|r_n\rangle
-
\sum_{k=0}^{n}
|q_k\rangle
\langle q_k|r_n\rangle
\end{aligned}
\label{eq:reorthogonalization}
\end{equation}
The recursion is terminated when the spectrum is converged under an increase of
the retained order according to a predefined normalised spectral-difference tolerance. The
specific tolerance, maximum order, broadening width, and reorthogonalization policy
used for each benchmark should be reported together with the corresponding
results.

\subsection{Vibrational Augmentation}
\label{sec:vibronic_extension}

\subsubsection{Electron--phonon Hamiltonian}

To incorporate vibronic structure, we augment the electronic Hamiltonian by
harmonic vibrational modes with linear electron--phonon coupling,\textsuperscript{\cite{giustinoElectronphononInteractionsFirst2017}}
\begin{equation}
\begin{aligned}
\hat H
&=
\hat H_{\mathrm{el}}
+
\sum_{l=1}^{M}
\omega_l\hat b_l^\dagger\hat b_l
+
\sum_{i=1}^{L}\sum_{l=1}^{M}
\lambda_{il}\,
\hat a_i^\dagger\hat a_i
(\hat b_l^\dagger+\hat b_l)
\end{aligned}
\label{eq:elph_hamiltonian}
\end{equation}

where
\(\hat b_l^\dagger,\hat b_l\) are bosonic creation and annihilation operators for
mode \(l\). We write
\(\lambda_{il}=g_{il}\omega_l\) when using the dimensionless coupling \(g_{il}\).
The model therefore retains multiphonon dressing generated by repeated action of
the linear coupling, while explicit quadratic or higher-order coupling terms and
anharmonic phonon potentials are not included. The off-diagonal electron-phonon coupling could be neglected as an option.\textsuperscript{\cite{lehrRoleVibronicCoupling2020}}

\noindent The product electron-phonon basis is
\begin{equation}
\begin{aligned}
|x,\mathbf m\rangle
&=
|x\rangle\otimes|\mathbf m\rangle,
\end{aligned}
\label{eq:product_basis}
\end{equation}

\noindent with fixed-electron-number occupation strings
\begin{equation}
\begin{aligned}
|x\rangle
&=
|x_1x_2\cdots x_L\rangle,
\quad
x_i\in\{0,1\},
\quad
\sum_i x_i=N_e
\end{aligned}
\label{eq:electronic_basis}
\end{equation}

\noindent and truncated local phonon states
\begin{equation}
\begin{aligned}
m_l&=0,1,\ldots,N_l^{\mathrm{full}}-1
\end{aligned}
\label{eq:phonon_basis}
\end{equation}

Once the coupled ground state is obtained, the same zero-temperature construction
of Sec.~\ref{sec:mkct_zero_temperature} is applied without modification.

\subsubsection{Huang-Rhys Selection of Phonon Modes}
\label{subsubsec: HR_selection}
Weakly coupled modes are discarded using a relative
Huang-Rhys criterion.  For mode $\alpha$, we define an effective scalar coupling
$g_{\alpha}$ and calculate Huang-Rhys factor as \textsuperscript{\cite{weiPolaritonicHuangRhys2023}}
\begin{equation}
  S_{\alpha} = \frac{g_{\alpha}^{2}}{2\omega_{\alpha}^{3}}
\end{equation}
Modes satisfying
\begin{equation}
  \frac{S_{\alpha}}{S_{\max}} \geq \eta,
  \qquad
  S_{\max}=\max_{\beta} S_{\beta}
\end{equation}
are retained. Unless otherwise stated, $\eta$ is set to $10^{-2}$.

\subsubsection{Compressed phonon representation}
\label{subsubsec: phonon_encoding}
To mitigate the rapid growth of the electron–phonon Hilbert space, each Huang–Rhys-selected phonon mode is compressed from a local Fock basis of dimension $d_{\mathrm{full}}$ to an encoded local basis of dimension $d_{\mathrm{enc}}$ for scalability.\textsuperscript{\cite{liEfficientQuantumSimulation2023}}
In the current workflow $d_{\mathrm{full}}=8$ and $d_{\mathrm{enc}}=4$ unless
overridden.  For each mode $\alpha$ we construct an encoder
\begin{equation}
  U_{\alpha}\in \mathbb{C}^{d_{\mathrm{full}}\times d_{\mathrm{enc}}},
  \qquad
  U_{\alpha}^{\dagger}U_{\alpha}=I
\end{equation}

Motivated by the half-encoded ground-state strategy of Li \textit{et al.},\textsuperscript{\cite{liEfficientQuantumSimulation2023}} we optimize each local encoder by extracting the dominant Schmidt subspace from a singular-value decomposition of the mode--environment coefficient matrix of the electron--phonon ground state. A complete macroiteration consists of one sweep over all retained phonon modes, as described in the surrounding text.

\subsection{Neural Quantum State Treatment of the Ground-State Structure}
\label{sec:nqs}
We represent the many-body ground state variationally using an autoregressive NQS
adapted from our earlier kernel-polynomial NQS framework.\textsuperscript{\cite{liuAbsorptionSpectraKernel2025}}
Both the electronic and electron-phonon implementations factorize the complex wavefunction into normalized conditional amplitudes and conditional phases, but use different local variables. The electronic model combines the two spin orbitals of each spatial orbital into a four-state variable, whereas the electron-phonon model appends binary variables for the encoded phonon qubits.

\subsubsection{Qubit Representation of Observables}
The electronic Hamiltonian is first mapped to qubits by the Jordan--Wigner
transformation,
\begin{equation}
\begin{aligned}
\hat a_j^\dagger
&=
\left(
\prod_{i=0}^{j-1} Z_i
\right)
\frac{X_j-iY_j}{2}
\\
\hat a_j
&=
\left(
\prod_{i=0}^{j-1} Z_i
\right)
\frac{X_j+iY_j}{2}
\end{aligned}
\label{eq:jordan_wigner}
\end{equation}
giving the Pauli expansion
\begin{equation}
\begin{aligned}
\hat H_q
&=
\sum_{k=1}^{K}
c_k P_k,
\quad
P_k=
\bigotimes_j \sigma_j^{(k)},
\quad
\sigma_j^{(k)}\in\{I,X,Y,Z\}
\end{aligned}
\label{eq:pauli_hamiltonian}
\end{equation}

Each retained vibrational mode $\alpha$ is truncated to a local Fock space of
dimension $d_{\alpha}=2^{q_{\alpha}}$. A local phonon state can therefore be
mapped exactly to a register of $q_{\alpha}$ qubits without introducing unused
computational-basis states. In particular, its occupation number $n_{\alpha}$ is
encoded as
\begin{equation}
n_{\alpha}
=
\sum_{j=0}^{q_{\alpha}-1}
2^{q_{\alpha}-1-j} b_{\alpha j},
\qquad
b_{\alpha j}\in\{0,1\}
\label{eqn: phonon_occ_map}
\end{equation}
with the state correspondence
\begin{equation}
|n_{\alpha}\rangle_{\mathrm{ph}}
\longleftrightarrow
|b_{\alpha 0}b_{\alpha 1}\cdots b_{\alpha,q_{\alpha}-1}\rangle_{q}
\end{equation}
Any operator $\hat O_{\alpha}$ acting on this finite local space can then be
expanded in the complete Pauli-string basis,
\begin{equation}
\hat O_\alpha
=
\sum_{P\in\{I,X,Y,Z\}^{\otimes q_{\alpha}}}
c_{P}P
\end{equation}
Because these Pauli strings are orthogonal under the Hilbert--Schmidt inner
product,
\begin{equation}
\operatorname{Tr}
\left(P^{\dagger}Q\right)
=
d_{\alpha}\delta_{PQ}
\end{equation}
the expansion coefficients are
\begin{equation}
c_{P}
=
\frac{1}{d_{\alpha}}
\operatorname{Tr}
\left(P^{\dagger}O_{\alpha}\right)
\end{equation}

\subsubsection{Complex Autoregressive Architecture}

For an ordered sequence of local variables $x=(x_1,\ldots,x_L)$, the NQS has
the polar form
\begin{equation}
\psi_{\theta}(x)
=
\exp\!\left[
\frac{1}{2}\sum_{l=1}^{L}\log p_{\theta}
\!\left(x_l\middle|x_{<l}\right)
+i\sum_{l=1}^{L}\varphi_{\theta,l}
\!\left(x_l\middle|x_{<l}\right)
\right]
\label{eq:nqs_autoregressive}
\end{equation}
Each conditional block receives a one-hot encoding of the preceding variables.
Separate feed-forward networks with rectified linear unit activations generate
the amplitude logits and phase values. Applying a log-softmax over the allowed
values of $x_l$ gives $\log p_{\theta}(x_l|x_{<l})$. Thus,
$|\psi_{\theta}(x)|^2=\prod_l p_{\theta}(x_l|x_{<l})$ is normalized by
construction, allowing the electronic implementation to sample configurations
sequentially without a Markov chain.

In the electronic NQS, $x_l$ specifies the joint occupation of the $\alpha$ and
$\beta$ spin orbitals in spatial orbital $l$ and takes one of the four values
$\{00,10,01,11\}$. Before each conditional distribution is normalized, the
model masks outcomes that cannot satisfy the prescribed numbers of $\alpha$ and
$\beta$ electrons. The wavefunction is therefore confined to the selected
fixed-particle-number sector. The implementation can share the amplitude and
phase blocks and impose spin-exchange symmetry; the calculations reported here
use shallow orbital-wise conditional networks.

For the encoded electron--phonon Hamiltonian, the electronic occupation bits and
phonon-qubit bits form a heterogeneous autoregressive sequence. For a variable
$x_l$ of local dimension $d_l$, the model concatenates the one-hot encodings of
$x_{<l}$ and returns $d_l$ amplitude and phase values. Here, $d_l=2$ for every
electronic or encoded phonon variable. The fixed-electron-number constraint acts
only on the electronic variables; the encoded phonon register spans the retained
local subspace without an additional occupation constraint. The electron--phonon
VMC routine evaluates the NQS probabilities over the enumerated restricted
encoded basis and samples from the resulting distribution.

\subsubsection{Variational Monte Carlo and Hybrid Sampled-IRL Training}

The autoregressive NQS approximates the ground state of the encoded electronic
or electron--phonon Hamiltonian as
\begin{equation}
\Psi_\theta(x)=\sum \psi_\theta(x)|x\rangle \qquad
  \psi_{\theta}(x)
  =\langle x|\Psi_{\theta}\rangle
\end{equation}
where $x=(n_{\mathrm{el}},n_{\mathrm{ph}})$ combines an electronic configuration
in the active sector with the encoded phonon-qubit labels. We minimize the
variational energy
\begin{equation}
  E_{\theta}
  =
  \frac{\langle \Psi_{\theta}|\hat{H}_{\mathrm{el-ph}}|\Psi_{\theta}\rangle}
       {\langle \Psi_{\theta}|\Psi_{\theta}\rangle}
\end{equation}
using configurations sampled from $|\psi_{\theta}(x)|^2$ and the corresponding
local energies,
\begin{equation}
E_{\mathrm{loc}}(x)
=
\sum_{x'}H_{xx'}\frac{\psi_{\theta}(x')}{\psi_{\theta}(x)}
\label{eq:nqs_local_energy}
\end{equation}
Stochastic VMC updates are interleaved with a sampled Krylov correction inspired
by the IRL optimization strategy of Liu \textit{et al.}\textsuperscript{\cite{liuImplicitlyRestartedLanczosEnables2026}}
We refer to this procedure as hybrid sampled-IRL. The hybrid scheme retains the stochastic parameter updates and periodically constructs a compact Ritz target on configurations sampled from the current NQS, which is then distilled into the network. This differs from applying the parameter-space IRL solver at every optimization step.

For sampled configurations $\{x_s\}$ with normalized multiplicity weights $w_s$, the local energies are centred and scaled as
\begin{equation}
y_s=\frac{E_{\mathrm{loc}}(x_s)-\sum_r w_rE_{\mathrm{loc}}(x_r)}
{\left[\sum_r w_r\left|E_{\mathrm{loc}}(x_r)-
\sum_jw_jE_{\mathrm{loc}}(x_j)\right|^2\right]^{1/2}}
\label{eq:sampled_irl_coordinate}
\end{equation}
A low-dimensional basis is then obtained by weighted Gram--Schmidt
orthogonalization of $\{1,y,y^2,\ldots,y^{K-1}\}$. If $B_{sk}$ denotes this
sampled basis, the overlap and projected Hamiltonian matrices are
\begin{equation}
\begin{aligned}
S_{kk'}&=\sum_s w_s B_{sk}^{*}B_{sk'}\\
H_{kk'}&=\sum_s w_s B_{sk}^{*}E_{\mathrm{loc}}(x_s)B_{sk'}
\end{aligned}
\label{eq:sampled_irl_matrices}
\end{equation}
We remove near-null eigenmodes of $S$ using a relative overlap threshold and
solve the resulting Hermitian generalized eigenproblem,
$H\bm c=E_{\mathrm R}S\bm c$, for the lowest Ritz pair. The ratio
$r_s=\sum_kB_{sk}c_k$ defines the configuration-space target
\begin{equation}
\log\psi_{\mathrm{target}}(x_s)
=
\log\psi_{\theta}(x_s)+\log r_s
\label{eq:sampled_irl_target}
\end{equation}
up to a global complex offset. We remove this offset by subtracting the
sample mean of the difference between the predicted and target values of
$\log\psi$, then minimize the corresponding mean-squared error using auxiliary
Adam steps with gradient-norm clipping. If the sampled Krylov basis collapses or
its overlap matrix is numerically singular, the correction is omitted for that
cycle and stochastic VMC training continues. The trained NQS provides the
normalized ground-state amplitudes and phases required by the subsequent
wavefunction-level MKCT recursion.

\section{Results and Discussions}
\subsection{MKCT Iterations and Numerical Features}
We first verified the reliability of the CF-MKCT method for producing absorption spectra by evaluating the convergence and accuracy of the MKCT workflow together with the number of iterations required to produce reliable spectra for simple molecules (\ce{NH3}, \ce{H2O}, \ce{C2}, \ce{N2}, \ce{O2}). The resulting spectra agree with FCI results when using spectral convergence criteria of $10^{-3}$ or tighter. In this case, reorthogonalisation was performed at every Lanczos iteration. The number of iterations needed generally varies with molecule and convergence criteria (\textbf{Table \ref{tab:NoM_conv}}). Systems with more active electrons and active orbitals do not always demand more iterations for accurate spectral reconstruction. The \ce{NH3} ($C_{3v}$) system has higher symmetry complexity, resulting in a more demanding number of iterations. Systems such as \ce{C2} require more iterations due to their multi-reference electronic structure.

\begin{figure}
    \centering
    \includegraphics[width=\linewidth]{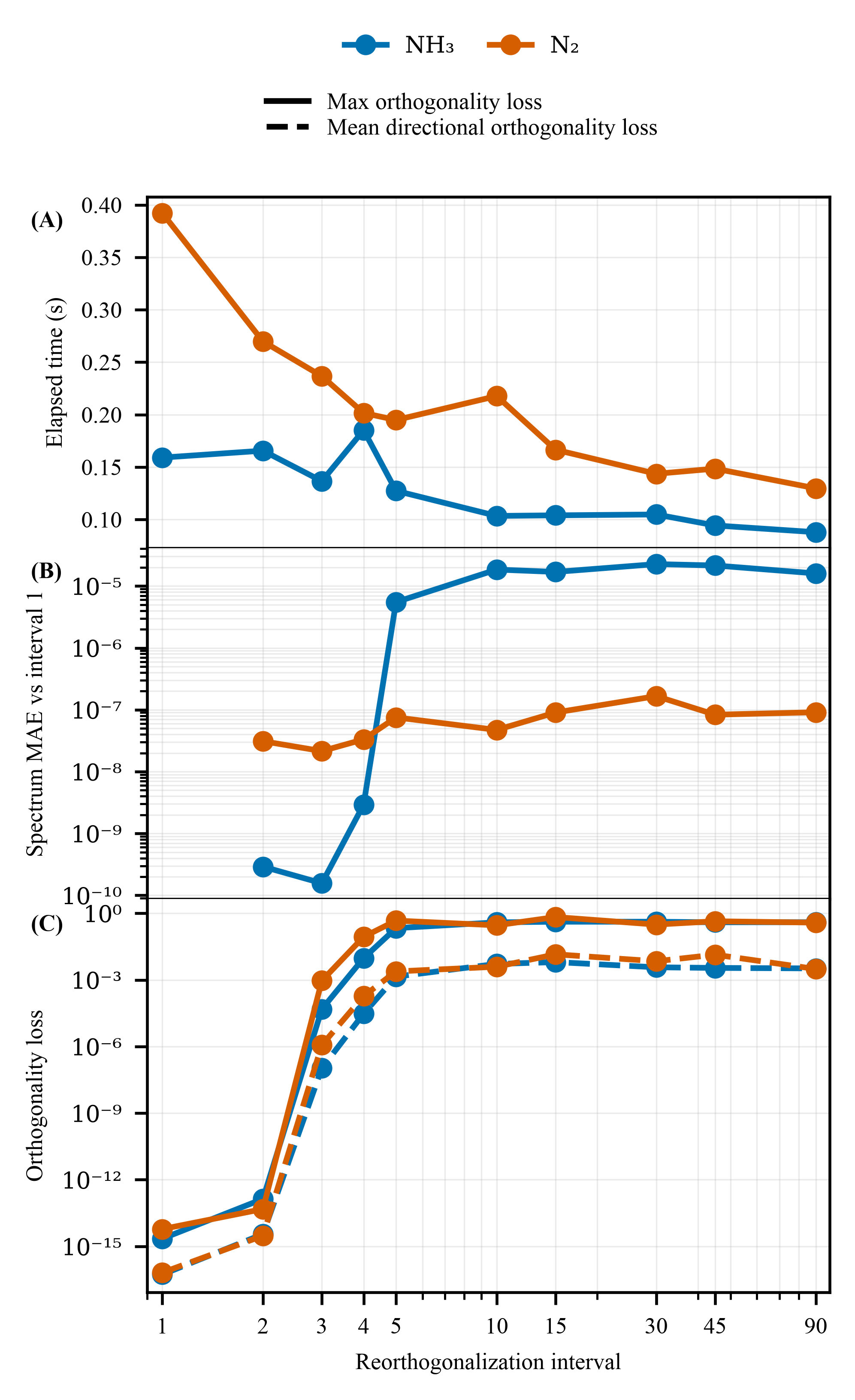}
    \caption{Reorthogonalization-interval benchmark for \ce{NH3} and \ce{N2}. (A) Elapsed time as a function of the reorthogonalization interval. (B) Mean absolute spectrum deviation relative to interval 1. (C) Maximum orthogonality loss and mean directional orthogonality loss versus reorthogonalization interval.}
    \label{fig:reortho}
\end{figure}

\begin{table}[h]
\centering
\caption{Summary of the number of iterations required to reach the target convergence}
\begin{tabular}{lccccc}
\hline
Molecule & $10^{-2}$ & $10^{-3}$ & $10^{-4}$ & $10^{-5}$ & $10^{-6}$\\
\hline
\ce{H2O} (8e, 6o)      & 50 & 50  & 70  & 110  & 110\\
\ce{NH3} (8e, 7o)     & 50 & 50  & 90  & 130  & 170\\
\ce{C2} (8e, 8o)      & 30 & 110 & 130  & 190  & 250\\
\ce{N2} (10e, 8o)      & 50 & 50 & 70 & 90  & 110\\
\ce{O2} (12e, 8o)      & 30 & 50 & 50 & 70  & 90\\
\hline
\end{tabular}
\label{tab:NoM_conv}
\end{table}
We then investigated the effect of reorthogonalisation during Lanczos iterations, as described in Section \ref{sec: cf-spectral-gen}. Although reorthogonalisation improves numerical accuracy per iteration, frequent reorthogonalisation reduces computational efficiency. We benchmarked the spectral prediction performance of the workflow on \ce{NH3} ($C_{3v}$) and \ce{N2} ($D_{\infty h}$). According to the results in \textbf{Figure \ref{fig:reortho}}, it is advisable to reorthogonalise the basis set at least once every 3 Lanczos iterations in order to obtain stable results. 
\subsection{NQS-driven MKCT Iterations (MKCT-NQS)}
\begin{figure}
    \centering
    \includegraphics[width=\linewidth]{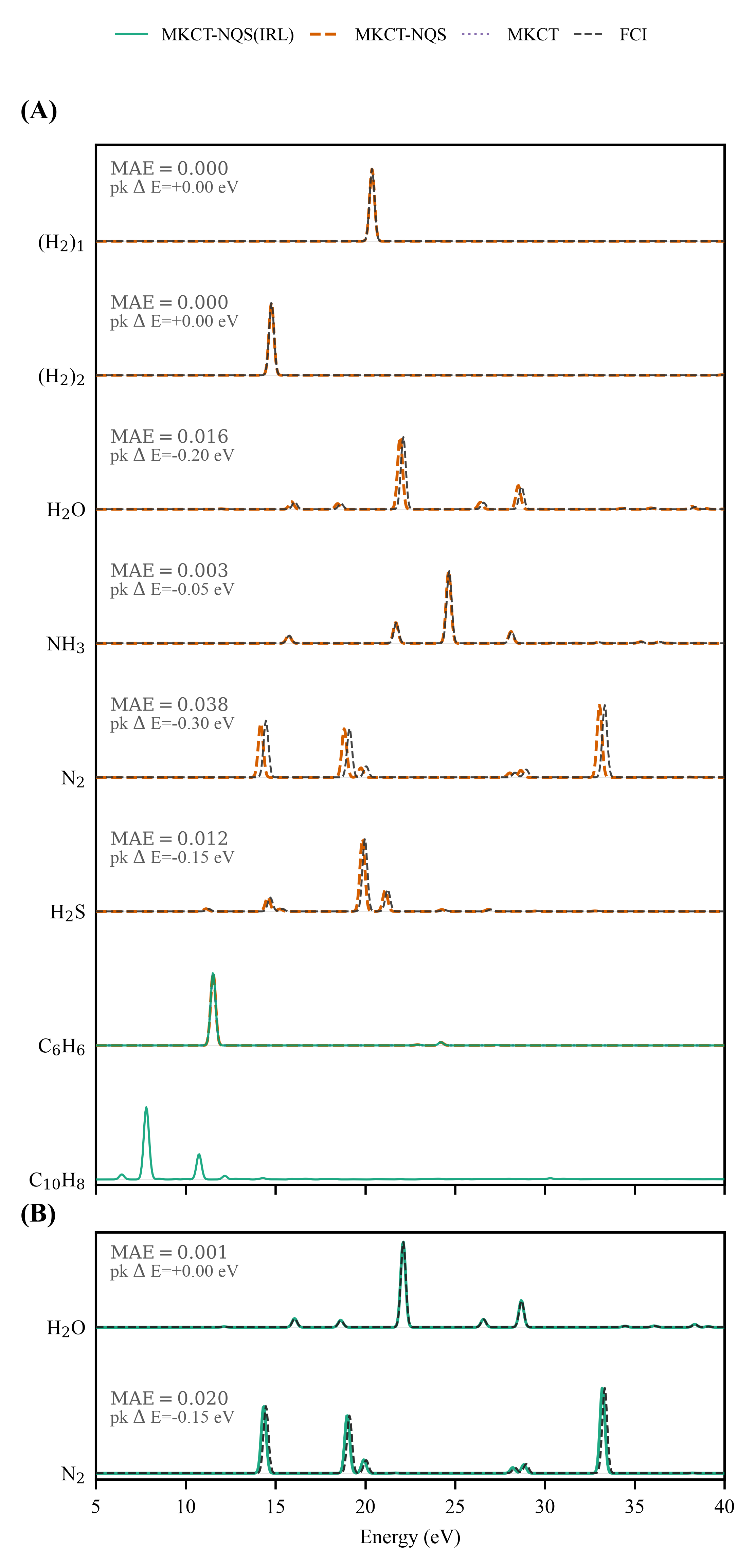}
    \caption{Electronic absorption spectra generated using MKCT, MKCT-NQS, MKCT-NQS(IRL) and FCI workflow. MAE was computed against FCI, peak difference (pk $\Delta E$) was measured between the strongest peak. Number of iterations required for (MKCT/MKCT-NQS/MKCT-NQS(IRL)): \ce{H2} (4/4/N.A.), \ce{(H2)2} (30/36/N.A.) \ce{H2O} (110/150/N.A.) \ce{NH3} (170/170/N.A.) \ce{N2} (110/310/N.A.) \ce{H2S} (130/150/N.A.) \ce{C6H6} (50/50/70) \ce{C10H8} (N.A./N.A./690)}
    \label{fig: MKCT_ea}
\end{figure}

\begin{figure}
    \centering
    \includegraphics[width=\linewidth]{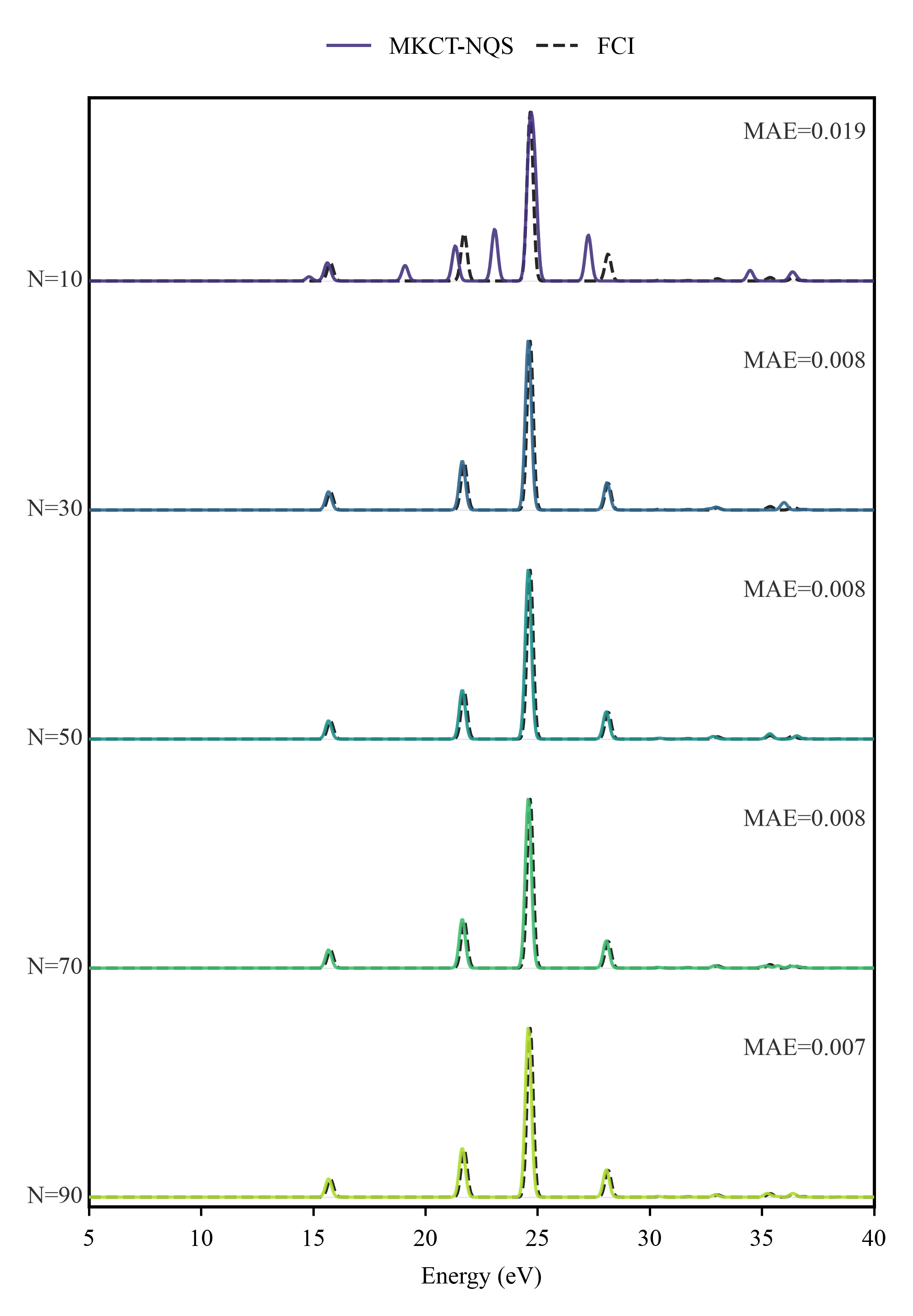}
    \caption{MKCT-NQS Convergence Plot of Spectral Generation. MAE values were calculated against the FCI spectrum of \ce{NH3}}
    \label{fig:NH3_conv}
\end{figure}

We then evaluated the performance of NQS-driven spectral prediction by benchmarking the resulting absorption spectra against FCI references and conventional MKCT calculations. All calculations were performed in the active spaces defined in Table \ref{tab:active_space}, with the frozen-core approximation. Quantitative validation was first performed for tractable molecular systems, indicating tight agreement between FCI, MKCT, and MKCT-NQS methods (\textbf{Figure \ref{fig: MKCT_ea}A}). We demonstrated that the baseline MKCT-NQS workflow can reach good agreement with FCI and MKCT references where available. For challenging cases such as \ce{H2O} and \ce{N2}, spectral quality can be improved by integrating implicit restarted Lanczos (IRL) into the NQS workflow (\textbf{Figure \ref{fig: MKCT_ea}b}). Having established agreement in these benchmark regimes, we further applied the same workflow to progressively larger molecular systems, including benzene, naphthalene, and anthracene. A convergence benchmark was also performed on \ce{NH3} (\textbf{Figure \ref{fig:NH3_conv}}), indicating that comparable accuracy can be reached with a convergence criterion of $10^{-3}$.

\begin{figure}[h]
    \centering
    \includegraphics[width=\linewidth]{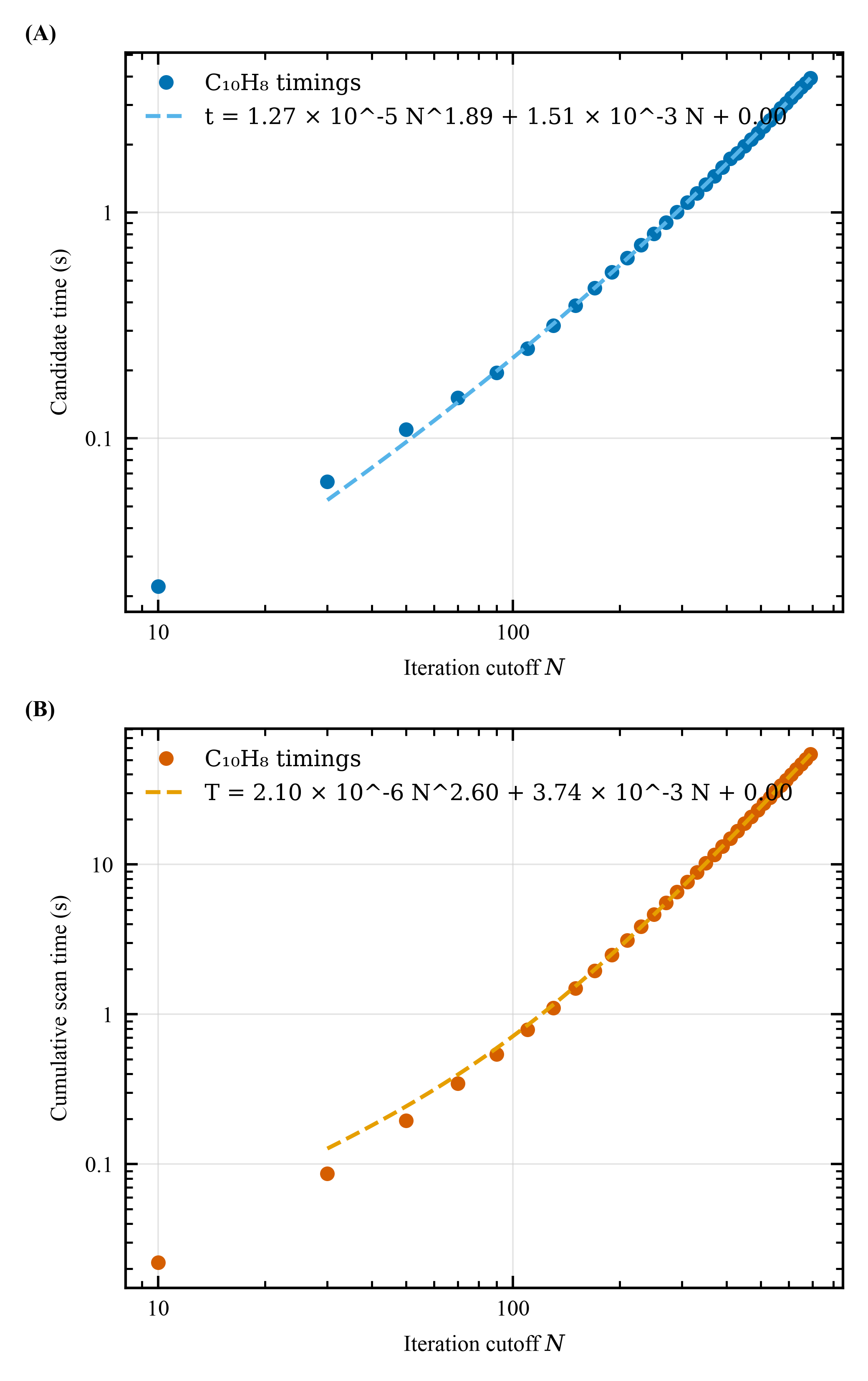}
    \caption{\ce{C10H8} electronic MKCT-NQS(IRL) cutoff-scan timing. Candidate time denotes the wall time for one normalized spectrum at a tested cutoff, including iterative construction and spectral reconstruction. Cumulative scan time sums the restarted candidate calculations across the automatic cutoff scan. The empirical fits were performed for $N\geq70$ using $aN^{p}+bN+c$}
    \label{fig:mkctnqsirl_profile}
\end{figure}

The computational profile of the electronic MKCT-NQS(IRL) workflow was examined using \ce{C10H8} as a representative large electronic benchmark. The automatic cutoff scan selected $N=690$ under a stringent spectral convergence threshold of $10^{-6}$, with a final candidate-spectrum time of 3.93 s and a total restarted cutoff-scan time of 54.84 s on the CuPy backend. These timings cover the cutoff-dependent spectral-generation stage and do not include the separate ground-state energy evaluation used to set $E_0$. To separate effective Krylov growth from fixed-grid spectral reconstruction and setup overhead, the timings for $N\geq70$ were fitted to $t(N)=aN^{p}+bN+c$ (\textbf{Figure \ref{fig:mkctnqsirl_profile}}). 

In each candidate spectrum profile (\textbf{Figure \ref{fig:mkctnqsirl_profile}A}), the non-linear Krylov-growth term becomes larger than the linear fixed-grid term only beyond $N\simeq215$, showing that the high-order contribution remains limited across the lower-cutoff regime traversed by the adaptive scan. Even under the stringent $10^{-6}$ criterion and full reorthogonalization, the selected \ce{C10H8} spectrum required only 690 recursions and the cutoff-dependent spectral-generation step took 3.93 s for the selected candidate. This recursion count is small compared with the size of the underlying electronic Hilbert space, and the Krylov-growth contribution remains practically contained while retaining direct spectral convergence control.

The cumulative profile (\textbf{Figure \ref{fig:mkctnqsirl_profile}B}) reflects the conservative restarted protocol used for this stored artifact. Each tested cutoff repeated earlier candidate construction, giving a total scan time of 54.84 s. Compatible current runs can reuse the Lanczos/Jacobi recurrence as the cutoff increases, converting repeated reconstruction into progressive refinement. In addition, spectral-shape comparisons generally tolerate thresholds closer to $10^{-3}$, which can reduce the selected cutoff relative to the stress-test value used here. Together, compact NQS-supported sampling, adaptive Krylov truncation, and matrix-free operator application make MKCT-NQS(IRL) an efficient route for large active-space electronic spectra.

\subsection{Vibrational Augmentation of MKCT-NQS Workflow}
We then extended our MKCT-NQS approach to model the electron-phonon coupled spectra. To resolve the widely appreciated difficulty of large Fock-space, we proposed
\begin{itemize}
    \item Phonon mode selection using Huang-Rhys factor (Sec.\ref{subsubsec: HR_selection}) 
    \item Two-level phonon-encoding scheme, primarily \textit{via} a singular value decomposition (SVD) based encoder, followed by qubit encoding (Sec.\ref{subsubsec: phonon_encoding})
\end{itemize}
IRL strategy was implemented to NQS optimisation considering the complexity of potential energy landscape for the electron-phonon coupled system. 

\begin{figure}[h]
    \centering
    \includegraphics[width=\linewidth]{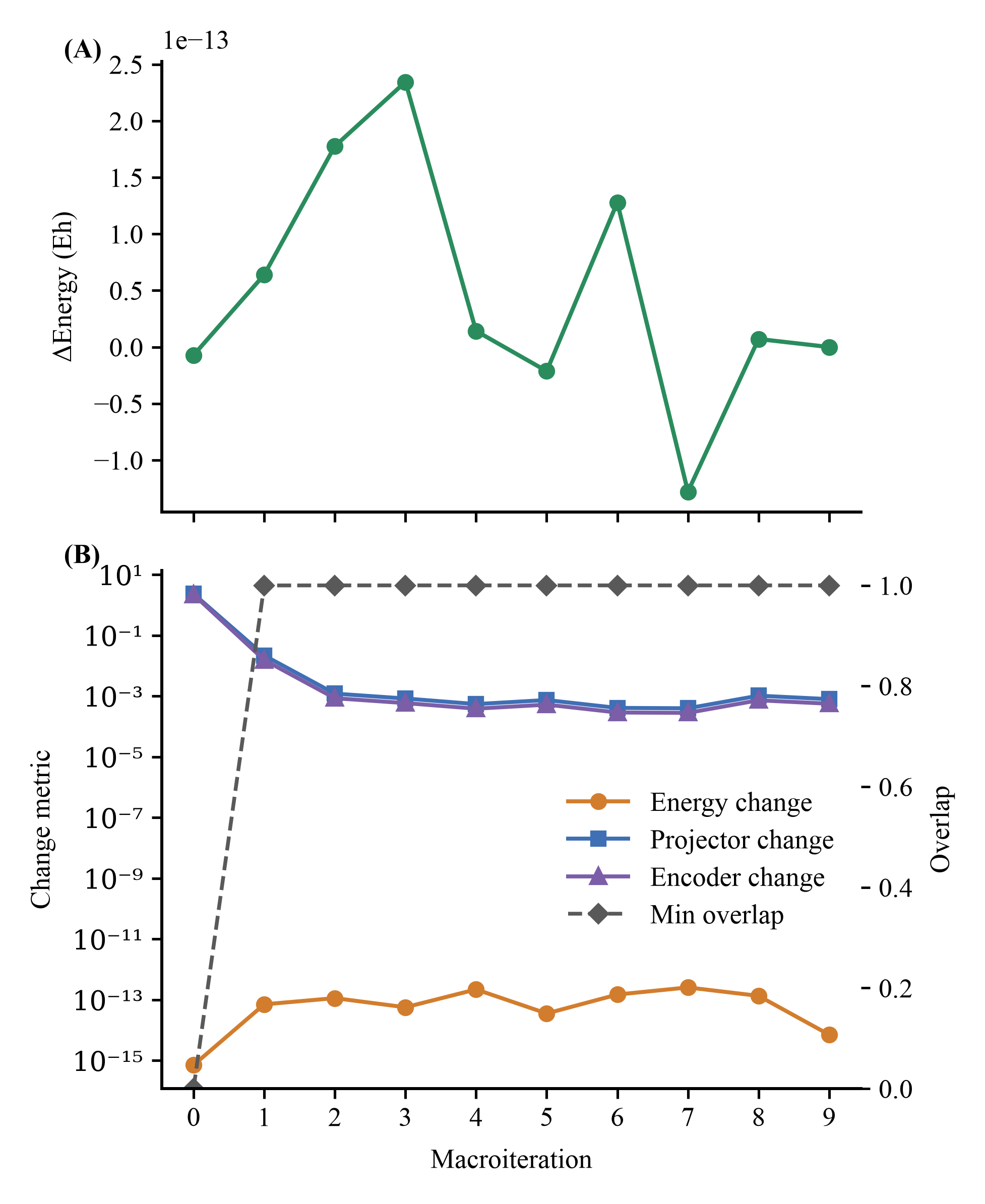}
    \caption{Convergence metrics of SVD for \ce{NH3} per macroiteration.}
    \label{fig: svd_conv_NH3}
\end{figure}

We first validated the performance of the SVD encoding. In practice, the initial low-occupation phonon basis already provides a physical starting point for the Huang-Rhys-selected modes, and a single sweep replaces each local basis by its dominant Schmidt subspace. Consequently, the fully encoded ground-state energy typically changes only weakly per macroiteration and the SVD discarded weight remains low (\textbf{Figure \ref{fig: svd_conv_NH3}}), while other metrics continue to relax. Therefore, the encoded energy is considered as a secondary diagnostic which is optional to be calculated.

We then benchmarked the integrated performance of this SVD-based encoding strategy and neural quantum state representation of ELPH structure using \ce{H2}, where exact diagonalization of ELPH coupled hamiltonian was plausible. A local phonon truncation of $n_{\max}=7$ was used, corresponding to a local Hilbert-space dimension of $d_\text{ph}=8$. SVD-based encoder compress the encoded Hilbert-space dimension to 4, which can be represented using 2 qubits. 

\begin{figure}[h]
    \centering
    \includegraphics[width=\linewidth]{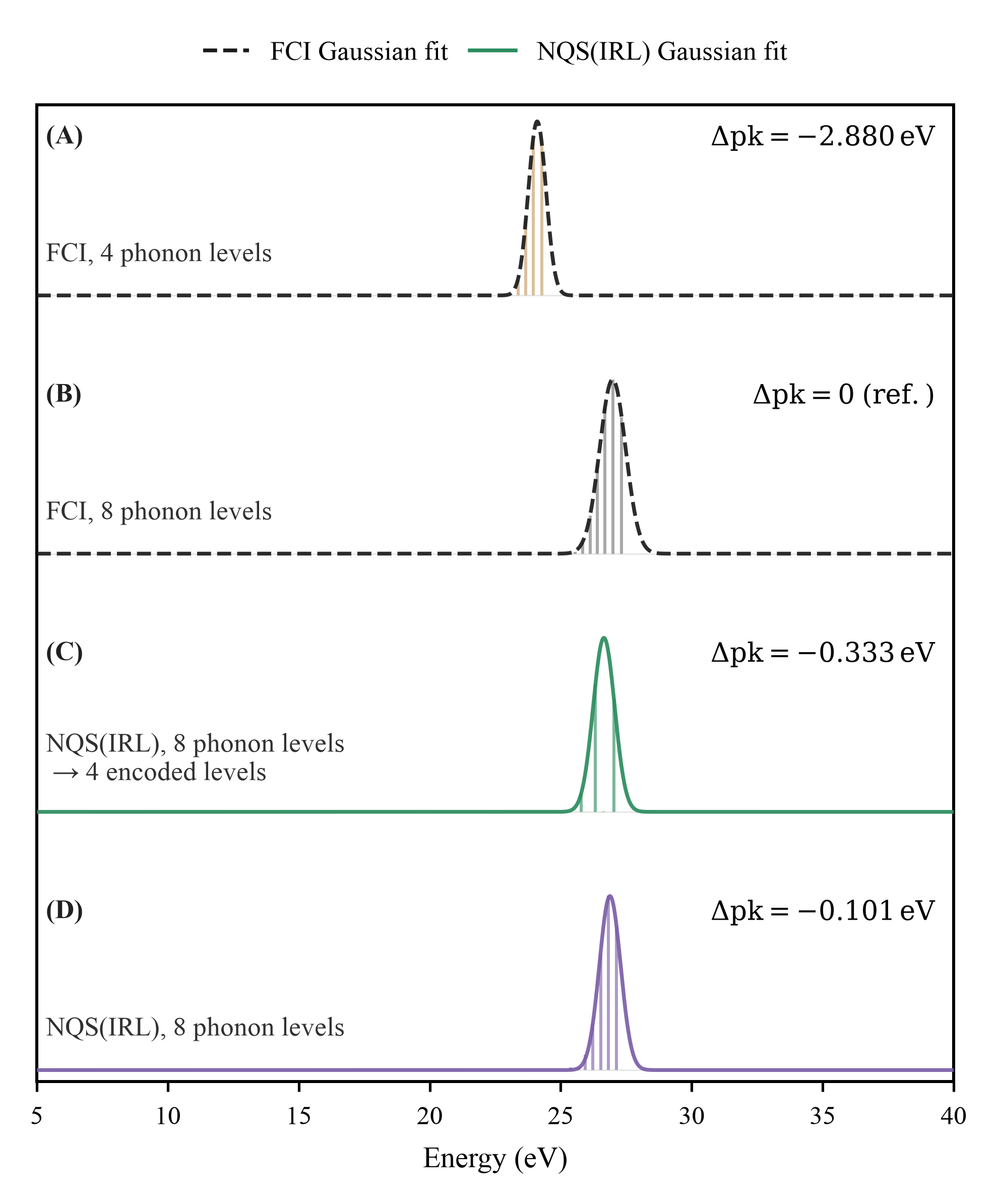}
    \caption{Electron-phonon coupled absorption spectra of \ce{H2} calculated using (A) FCI spectrum, maximum phonon occupation truncated to 4 phonon levels (B) FCI spectrum, maximum phonon occupation truncated to 8 phonon levels (C) MKCT-NQS-ELPH with implementation of SVD encoding (D) MKCT-NQS-ELPH without SVD encoding.}
    \label{fig:H2_elph}
\end{figure}

Although the low-rank encoded basis reduces the number of resolved vibronic sidebands, these sidebands would typically remain unresolved in practice. We therefore compare the dominant spectral envelope by fitting the local peak maxima to a single Gaussian. The spectrum (\textbf{Figure \ref{fig:H2_elph}C}) obtained with the SVD-encoded phonon basis shows comparable peak-position agreement with the FCI reference than the non-encoded result(\textbf{Figure \ref{fig:H2_elph}D}). At the same time, the envelope RMSE indicates that the two fitted spectra have comparable overall accuracy. For comparison, reducing the bare phonon space to four levels shifts the FCI peak by $-2.88$ eV relative to the 8-level reference (\textbf{Figure \ref{fig:H2_elph}A/B}). The SVD-based encoder, by contrast, recovers the dominant peak to within $-0.337$ eV of the 8-level reference while using a compact local representation. This substantial recovery in spectral accuracy shows that the encoder retains the spectroscopically relevant vibronic subspace rather than merely reducing the Hilbert-space dimension.

\begin{figure}[h]
    \centering
    \includegraphics[width=\linewidth]{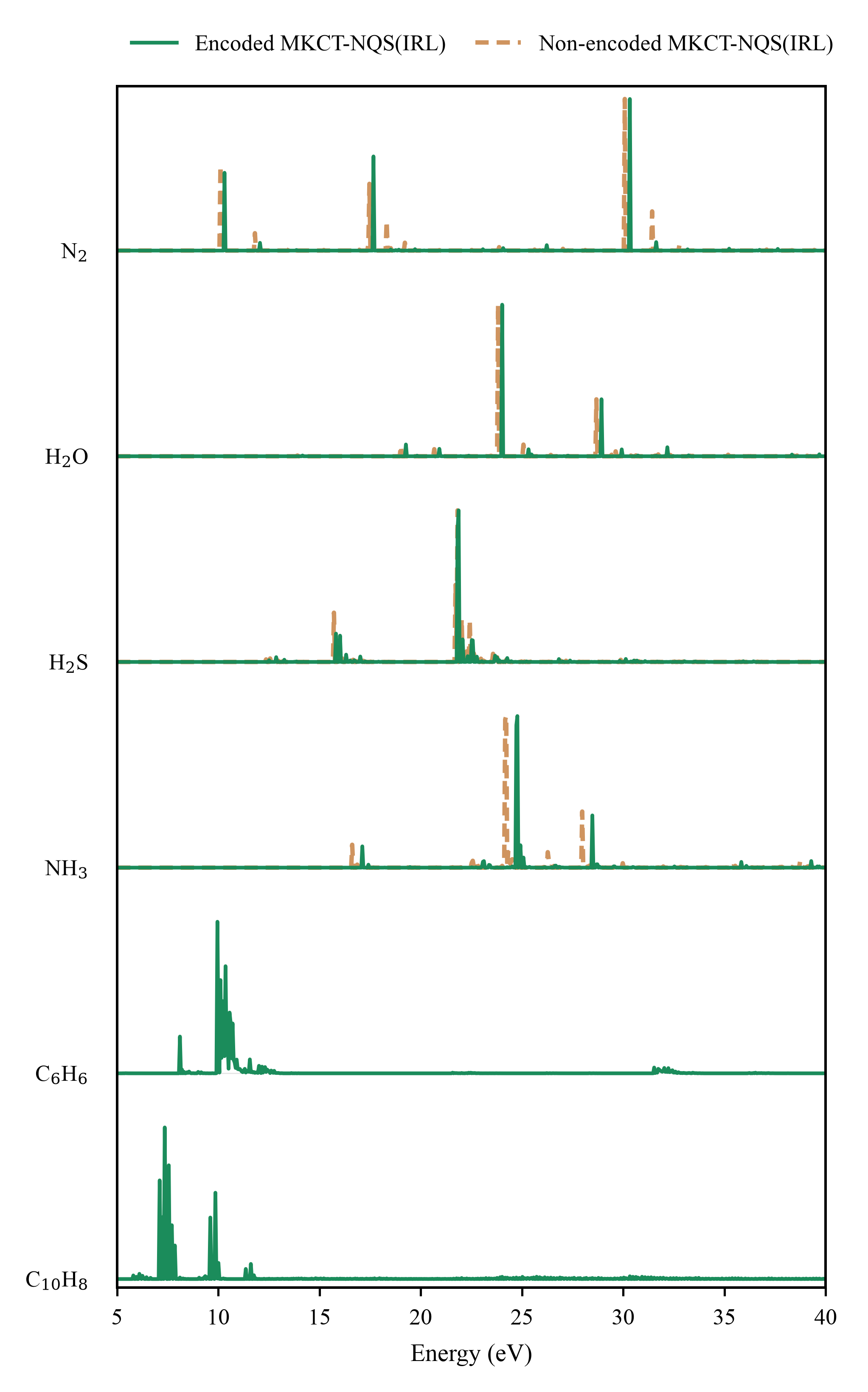}
    \caption{Electron-phonon coupled absorption spectra obtained with MKCT-NQS(IRL). Number of iteration required (encoded/non-encoded) per spectra \ce{N2} (1990/150), \ce{H2O} (970/150), \ce{H2S} (1650/550),  \ce{NH3} (5850/910), \ce{C6H6} (3070/N.A.), \ce{C10H8} (1050/N.A.).  specified in the text. Each spectrum is normalized independently.}
    \label{fig:elph_aromatic_stack}
\end{figure}

Having examined the performance of the SVD-based phonon encoding in the \ce{H2} benchmark, we next applied the same electron-phonon workflow to other molecular systems. Figure~\ref{fig:elph_aromatic_stack} compares encoded and non-encoded MKCT-NQS(IRL) spectra for \ce{N2}, \ce{H2O}, \ce{H2S}, and \ce{NH3}. In each comparison, both representations use a maximum local phonon occupation of $n_\mathrm{max}=7$, a spectral convergence criteria of $10^{-6}$, and a Huang-Rhys mode-selection threshold of $S/S_\mathrm{max}\geq0.01$. The two representations recover similar dominant spectral envelopes across all four molecules, showing that the compressed phonon basis retains the principal spectral structure produced by the corresponding non-encoded NQS calculation. Differences remain in individual sideband positions and intensities. As both curves are generated from independently trained NQS, this comparison measures the combined effect of phonon representation, NQS training, and spectral construction. We then progressively demonstrated the ability of such workflow in modelling more complexed aromatic molecules, benzene (\ce{C6H6}) and Naphthalene (\ce{C10H8}). Due to limitation of computational resource, the calculations use Huang-Rhys thresholds of $S/S_\mathrm{max}\geq0.10$ and $S/S_\mathrm{max}\geq0.20$, and spectral convergence thresholds of $10^{-4}$ and $10^{-3}$ correspondingly.

\section{Conclusion \& Outlook}
We introduced Aurora, a NQS ground-state-based framework for predicting molecular electronic absorption spectra without explicitly constructing an excited-state manifold. By combining variational many-body neural quantum ground state representations with zero-temperature memory-kernel coupling theory, Aurora reformulates spectroscopic response in terms of wavefunction-level operations on a single ground state. This provides a direct interface between scalable ground-state solvers, including neural quantum states, and frequency-domain spectroscopy.

We further demonstrated that this framework can be extended to include vibronic coupling through an explicit electron-phonon augmentation \textit{via} neural quantum state trained by variational monte-carlo. In this extension, Huang-Rhys factor based mode selection identifies the vibrational degrees of freedom most relevant to the spectrum, while local SVD-based phonon encoders reduce the cost of the retained phonon Hilbert spaces by exploiting their ground-state structure. The resulting compressed representation preserves the dominant vibronic spectral features while substantially reducing the dimensional overhead associated with explicit phonon degrees of freedom.

More broadly, Aurora separates spectral modelling into two components — an accurate and scalable representation of the ground state and a recursive response construction built upon it. The accuracy of the resulting spectra is therefore controlled at both levels, in addition to an overall benchmark against FCI spectra where computationally affordable. First, the quality of the ground-state representation is assessed independently through variational convergence and, where tractable, direct comparison with exact or high-accuracy reference energies and wavefunctions. Second, the recursive spectral reconstruction is systematically converged by increasing the recursion order until the resulting spectral profile satisfies a predefined numerical tolerance. This separation allows errors originating from the many-body ground state and from the response reconstruction to be diagnosed independently, while making the framework largely agnostic to the underlying ground-state ansatz. Importantly, this two-level architecture makes Aurora naturally extensible, as additional physical effects can be introduced at the level of the ground-state construction without reformulating the recursive response machinery, such as vibronic augmentation demonstrated here.

These results further demonstrate that molecular absorption spectroscopy need not be formulated primarily as an excited-state eigenvalue problem. Instead, spectroscopic information can be recovered directly from a ground-state many-body representation and its response to physically relevant operators. Aurora therefore provides a scalable and extensible route toward electronic spectroscopy, with vibronic augmentation as a systematic extension when required.
\section{Acknowledgement}
We thank financial support from the National Natural Science Foundation of China (Grant
Nos. 22273075 and 22361142829) and the Zhejiang Provincial Natural Science Foundation (Grant
No. XHD24B0301).  X.H. thanks Wei Liu for useful discussions.
\section{Author Contribution}
X.H. (Conceptualization, Methodology, Software, Validation, Investigation, Formal analysis, Visualization, Writing - original draft, Writing - review and editing). W.D. (Conceptualization, Methodology, Formal analysis, Investigation, Supervision, Project administration, Funding acquisition, Resource, Writing - review and editing). All authors approved the final manuscript.

\bibliography{Dou_Group}

\end{document}